# Modularity, Noise and natural selection

**Gabriel Marroig • Diogo A R Melo • Guilherme Garcia**

gmarroig@usp.br   diogro@usp.br   wgar@usp.br

Laboratório de Evolução de Mamíferos

Departamento de Genética e Biologia Evolutiva

Instituto de Biociências Universidade de São Paulo

Caixa Postal 11.461, CEP 05422   970, São Paulo   SP, Brasil.

Phone: +55 11 3091-8758, Fax: +55   11 3091-7553

**Abstract**

Most biological systems are formed by component parts that to some degree are inter-related. Groups of parts that are more associated among themselves and are relatively autonomous from others are called modules. One of the consequences of modularity is that biological systems usually present an unequal distribution of the genetic variation among variables. Estimating the covariance matrix that describes these systems is a difficult problem due to a number of factors such as poor sample sizes and measurement errors. We show that this problem will be exacerbated whenever matrix inversion is required, as in directional selection reconstruction analysis. We explore the consequences of varying degrees of modularity and signal-to-noise ratio on selection reconstruction. We then present and test the efficiency of available methods for controlling noise in matrix estimates. In our simulations, controlling matrices for noise vastly improves the reconstruction of selection gradients. We also perform an analysis of selection gradients reconstruction over a New World Monkeys skull database in order to illustrate the impact of noise on such analyses. Noise-controlled estimates render far more plausible interpretations that are in full agreement with previous results.

## 2. Introduction

The study of biological systems and its component parts, whether molecules, cells, tissues, organisms and its forming parts, and even species and their interactions, is rapidly converging to the central theme of modularity. This refers to the connections among some of the component parts of a biological system (genes or morphological traits, for example) and the lack of such associations among other parts of the same system. (Olson and Miller 1958, Berg 1960, Wagner et al. 2007). The notion that interacting parts are not independent is intuitive and appears early in the history of Biology (see Mayr 1982). Therefore, modularity is quickly becoming one of the central questions in modern biology (Wagner et al. 2007; Klingenberg 2008) and a point of convergence of various specialties and areas (Mathematics, Statistics, Genetics and Genomics, Evolutionary Biology, Ecology, Biochemistry and Physiology).

In biology, several types of modules have been recognized, including: a) functional, consisting of characters or features that act together on performing a task or function and are quasi-autonomous  in relation to other functional sets: b) developmental, which corresponds to parts of an embryo that are relatively autonomous with respect to pattern formation and differentiation, or an autonomous signaling cascade; c) variational, composed of characters that vary together and are relatively independent of other such sets (Wagner et al. 2007).

The study of modularity is centered on statistical estimation of association among traits (Olson and Miller 1958, Berg 1960), whether such association is measured by correlation, covariance, or distance/similarity measures, it is usually represented by matrices. Even if a particular system or network does not present a modular structure or is not being interpreted under this theory, associations among traits, parts, genes or lineages will still be quantified by statistical association or dissociation matrices among these elements. We will focus here on correlation or covariance matrices (from now on $C$-matrix) among variables, although the same problem appears in any statistics of association/dissociation among the component parts of any system. As biologists, what we usually do is to sample nature and infer properties from natural systems using measures and statistics obtained from such samples.  We should be aware of the fact that by sampling a population we do not have the true population parameter values but only estimates of these quantities. These estimates will be approximations that should converge to the true population value depending on a number of things, such as: sample size, number of parameters considered, precision of the measuring device and the quality of parameter estimators themselves, measured by their precision and accuracy (Sokal and Rohlf, 1995). The general trend is that, as the ratio between number of parameters and sample size decreases, signal-to-noise ratio will increase. For C-matrices this effect is summarized by their sampling distribution (the Wishart distribution) and can be expressed in terms of their eigenvalues (see next section and Meyer and Kirkpatrick, 2008).

In this paper we illustrate the problem of noise in matrix estimation using modularity as our framework and addressing effects of matrix estimates in the context of natural selection. The evolutionary response of a set of quantitative traits is described by $\boldsymbol{\Delta\bar{z}} = \boldsymbol{G\beta}$, where $\boldsymbol{\Delta\bar{z}}$ is the vector of differences in means between generations, $\boldsymbol{\beta}$ is the selection gradient vector, and $\boldsymbol{G}$ is the additive genetic covariance matrix (Lande 1979). Rearranging the evolutionary response equation, the pattern of selection responsible for populations' divergence can be reconstructed from observed mean differences using the relationship:

$$\boldsymbol{\beta} = \boldsymbol{G}^{-1}\boldsymbol{\Delta z} \qquad (1)$$

where $\boldsymbol{\beta}$ is the cumulative selection gradient summed over generations (net-$\boldsymbol{\beta}$ *sensu* Lande, 1979), $\boldsymbol{G}^{-1}$ is the inverse of $\boldsymbol{G}$ and $\boldsymbol{\Delta\bar{z}} = (\bar{z}_i - \bar{z}_j)$ is the difference in means between populations $i$ and $j$ (Lande 1979; Lofsvold 1988; Cheverud 1996). Selection reconstruction can be extremely useful in understanding patterns of multivariate selection within a microevolutionary context (Lande and Arnold 1983, Boag 1983; Grant and Grant 1995), and if certain assumptions hold (see Marroig and Cheverud 2001, 2004, 2005, 2010), can be extended to a macroevolutionary context.

We illustrate here how noise associated with sampling in matrix estimation can lead to error in $\boldsymbol{\beta}$ reconstruction using a simulation approach. We also illustrate how changing both number of modules and magnitude of association among elements in a system will affect matrix estimation and $\boldsymbol{\beta}$ reconstruction. We present possible solutions which could help ameliorate the noise problem in matrix estimation and test the performance of such solutions. Finally, we illustrate the analyses of selection reconstruction in a well studied case, New World Monkeys skull evolution, and show how previously estimated $\boldsymbol{\beta}$'s (Marroig and Cheverud 2005) are most likely dominated by noise and how estimates of these gradients, after controlling for noise, lend an interpretation in full agreement with results obtained using a different approach on the same dataset (Marroig and Cheverud 2010).

## 3 Materials & Method

### 3.1 Background - extension approach

The study of quantitative traits is largely based on $C$-matrices. Estimation of these matrices is usually done in a straightforward manner, simply sampling a population of interest and estimating the population covariance matrix for a set of $p$ traits using the sample covariance matrix obtained from a group of $n$ individuals. This method has its appeal derived from the fact that this estimator is the maximum likelihood (ML) estimator under multivariate normality.

However, when dealing with small datasets, in which the number of parameters being estimated is close to the number of individuals or even an order of magnitude smaller, one can expect substantial error to be present in the estimated matrix. This problem is even more severe when sample size $n$ is smaller than the number of traits $p$, when the sample matrix becomes singular. This is clearly a problem, which stems from the fact that the ML estimator is only asymptotically convergent to the true value, and for small datasets it is possible to improve greatly on it (Stein 1956). The effect can become even more critical in the genomic era where large matrices are estimated for hundreds or thousands of genes with very few observations. Usually in quantitative genetics the ML estimator is good enough, so this convergence problem is overlooked in most cases. But when theory demands matrix inversion, this "noise" associated with under-sampling becomes very apparent. In particular it greatly affects selection reconstruction analysis and must be taken into account in such studies, as we show later on.

One simple example of the noise involved in matrix estimation can be visualized in the complete absence of modularity (40 independent traits), so even when dealing with uncorrelated traits of equal variance (or with the correlation matrix for these traits) where one would expect eigenvalues of the $C$-matrix to be equal, in practice only with very large sample sizes do these eigenvalues tend to approach the expected value (see Figure 1a and Horn 1965). This phenomenon occurs because although ML estimator for $C$-matrices provides an unbiased estimator for the mean of its eigenvalues, as expected by the sampling distribution of *C*-matrices, their variance is overestimated (Meyer and Kirkpatrick 2008). This "spreads" individual eigenvalue estimates, increasing the larger and decreasing the smaller ones.

We can visualize the effect of noise by plotting the distribution of estimated correlations for different sample sizes, as presented in Figure 1b for this uniform matrix. Notice that, by chance alone, with small sample sizes, while the average of all 780 pairwise correlations is zero as expected, the range of observed values is between -0.6 and 0.6 for $n = 20$, and even with $n = 200$ we still observe correlations ranging from -0.25 to 0.25. All those correlations would indicate moderate association between traits despite the fact that they are independent. The situation is dramatically improved by using the extension method detailed below.

The extension approach was recently introduced by Hayden and Twede (2002) (also Twede and Hayden 2004) as way to improve performance of spectral filtering algorithms based on covariance matrix inverses. The idea is based on the argument that matrices are estimated with some degree of error and this noise is actually amplified and dominates inverted matrices. A *C*-matrix (**C**) can be represented by the following equation:

$$\boldsymbol{C} = \boldsymbol{V}^T \boldsymbol{\Lambda} \boldsymbol{V} \tag{2}$$

where $\boldsymbol{V}$ is a square matrix of normalized eigenvectors, $\boldsymbol{\Lambda}$ is a square diagonal matrix of eigenvalues $\lambda_i$ ; the superscript $^T$ denotes matrix transpose. The inverse of $\boldsymbol{C}$ is given by:

$$\boldsymbol{C}^{-1} = \boldsymbol{V}^T \boldsymbol{\Lambda}^{-1} \boldsymbol{V} \tag{3}$$

where the diagonal elements of $\boldsymbol{\Lambda}^{-1}$ are the inverse of each eigenvalue $\left(\frac{1}{\lambda_i}\right)$. When $\lambda_i$ becomes small, the estimate $\boldsymbol{C}^{-1}$ becomes a poor estimate of the true inverse. This is due to a number of factors, but particularly in biological systems because of sampling error in the estimation and the fact that modularity reduces the dimensionality of $\boldsymbol{C}$. The smallest eigenvalues are the ones that concentrate the most noise in the estimate of $\boldsymbol{C}$, and will become the dominant figure in the inverted matrix $\boldsymbol{C}^{-1}$, which is obvious if we think in terms of the inverted eigenvalues. For example, a 0.1 eigenvalue in $\boldsymbol{\Lambda}$ will become 10 in $\boldsymbol{\Lambda^{-1}}$, and a smaller one of 0.0001 will become 10000. Because those last eigenvalues are precisely those with larger error in their estimates (Figure 1) this unduly contribution to $\boldsymbol{C^{-1}}$ will affect the performance of analyses based on this matrix and consequently the inferences draw from it.

Here we treat the sequence of sorted eigenvalues as a discrete function defined on (say) the natural numbers from 1 to $p$, where $p$ is the number of traits. The second derivative of this function is defined by $\lambda_{i-1} - 2\lambda_i + \lambda_{i+1}$ for the $i$-th point. Using this definition it is possible to find the region of the eigenvalues distribution corresponding to noise by inspecting these values and finding the minimum value of the second derivative. Subsequent eigenvalues should then be replaced by the last reliable eigenvalue (the extended eigenvalue). This was the original criterion from Hayden and Twede (2002) to find the noise-floor region. We here modified this approach by taking the absolute values of this second derivative and then calculating the variance of those values in groups of 3 to 5 consecutive second derivatives. When this variance approaches zero (0.001 or 0.0001) the noise-floor region is reached and subsequent eigenvalues should be replaced by the last reliable eigenvalue (see Figure 2 for an example of both approaches). The point when the variance is sufficiently close to zero is somewhat variable, depending on the scale of the organism (the total amount of variation in the matrix) and the amount of noise in the estimated matrix, and should be analyzed graphically on a case by case basis. In an attempt to minimize this effect, instead of working directly with the eigenvalues, we use the amount of variation that each one explains, or, more clearly, we divide each eigenvalue by their sum. We compare both criteria in the following sections.

Notice that $\boldsymbol{C}$ (and $\boldsymbol{C^{-1}}$) are not truncated and still maintain their full rank, and that eigenvectors ($\boldsymbol{V}$) are not changed but only adjusted in their variances (eigenvalues) in dimensions with very little variation. This means that matrix structure is essentially preserved by using this approach.

### 3.2 Simulation and bootstrap

We used a simulation strategy with re-sampling via bootstrap to explore the impact of sampling error, modularity and noise on reconstructing selection gradients. This strategy was designed to represent the variation in modularity patterns and, most notably, magnitude that is observable across skull variation in mammalian taxa (see Marroig and Cheverud 2001; Porto et al. 2009; Oliveira et al. 2009). The first step was to create a series of *C*-matrices with varying degrees of modularity and overall integration. All matrices are 40x40 in rank and simulate morphological data. Notice that all these matrices represent biological systems with low dimensionality; in other such systems, like gene co-expression databases, for example, dimensionality can easily reach the order of thousands (e.g.: Oldham et al. 2006, Nowick et al. 2009). We start with uniform *C*-matrices, with equal eigenvalues, then move to matrices with two separate modules (A matrices), then four modules (B), eight modules (C) and finally matrices with eight modules and two sub-modules within each main module (D). In all these matrices all 40 traits were divided equally among modules. In each of these modular structures we also increase both between and within module correlation, which alters the influence of the first principal component (size) on the total amount of variation (matrices A through D of types 1, 2 and 3 and two uniform matrices – see Supporting Information). Using these *C*-matrices we created 14 multivariate normal distributions with null mean, and drew from each one populations with 10,000 individuals. These are meant to represent natural populations of interest.

Next, 1,000 bootstrap samples of 50, 100, 200 and 500 individuals were taken from populations, representing experiments using different sample sizes. Each of these individual samples was used to calculate a *C*-matrix estimate.

We are interested in how well these sample *C*-matrices can be used to estimate selection gradients. To determine this we generate a random known selection gradient ($\boldsymbol{\beta}_T$), which is used in Lande's equation along with the population *C*-matrix to determine the observed $\boldsymbol{\Delta}\bar{\mathbf{z}}$. Using this $\boldsymbol{\Delta}\bar{\mathbf{z}}$ and sampled *C*-matrices we use Equation 1 to calculate an estimate of the selection gradient ($\boldsymbol{\beta}_S$). The similarity between $\boldsymbol{\beta}_T$ and $\boldsymbol{\beta}_S$ is then a measure of the quality of *C*-matrix estimation and of its inverse. This procedure is summarized in Figure 3.

We use vector correlation to compare all vectors in this study. Using random vectors draw from a multivariate normal distribution with 40 elements we determined that any vector correlation with absolute value above 0.5 is significant at $p < 0.001$. The same sample $C$-matrices were then submitted to the extension approach and a number of comparisons were made to check if controlling noise improves the results. We present below three criteria for selecting the numbers of eigenvalues that should be retained: an optimal criterion, a variance criterion and a minimum criterion.

The optimal criterion (OC) is the best possible fit between $\boldsymbol{\beta}_T$ and the reconstructed selection gradient based on the noise-controlled sample $C$-matrix ($\boldsymbol{\beta}_{opt}$). For every $C$-matrix sampled from the population we determine how many retained eigenvalues gives the highest correlation between these selection gradients. The rationale here is to have a benchmark against which to compare the impact of the extension procedure, and is obviously not applicable to real world problems where we do not know the true population $C$-matrix or the true selection gradient.

We introduce here the variance criterion (VC), in which we calculate the cut-off estimate described in the background section without any consideration to true population values. It is entirely based on the decay of variance of the second derivative of sorted eigenvalues. The comparison between these reconstructed selection gradients based on the VC ($\boldsymbol{\beta}_{var}$) with those obtained from the OC will be useful to benchmark whether or not we are still able to find a reasonable cut-off point without any *a priori* knowledge of the population $C$-matrix.

The minimum criterion is the original cut-off estimative of Twede and Hayden (2004) for the extension method, which takes the minimum of the second derivative of sorted eigenvalues as the cut-off point.

We also apply another noise control technique described for covariance matrices estimation. The shrinkage method is based on a theorem by Stein (1956) that shows that it is often possible to improve the maximum likelihood (ML) estimator for the $C$-matrix of a given high-dimensionality data set. The idea is to use a target matrix and find the linear combination between the ML estimator and this arbitrary target matrix that best describes the true covariance structure of the population. A good choice for the target matrix in our case is a diagonal matrix of variances with null covariances. This particular target only modifies covariances, leaving variances untouched. Using a positive definite target matrix, this procedure guarantees a resulting matrix that is always positive definite and well conditioned. The optimal linear combination between thetarget and original matrix, called shrinkage intensity, is determined analytically as to minimize the mean squared error of the resulting shrinkage matrix estimate (Ledoit and Wolf 2003). This analytical calculation of shrinkage intensity is very robust, and a poor choice of target matrix will result in a linear combination that privileges the original matrix estimated directly from the data set. For a ML estimated C-matrix $\boldsymbol{C}$, the shrinkage estimate $\boldsymbol{C}^*$ is given by:

$$\boldsymbol{C}^* = \alpha\boldsymbol{T} + (1 - \alpha)\boldsymbol{C} \tag{4}$$

where $\boldsymbol{T}$ is the target matrix and $\boldsymbol{\alpha}$ is the shrinkage intensity. For an extensive description of the method, different shrinkage targets and how to calculate shrinkage intensity for such targets see Schafer and Strimmer (2005). The shrinkage method results were then compared to our cut-off estimative for the extension method in their efficiency in estimating the "true" $C$-matrix of the whole sample.

In all methods we also measure the amount of variation explained by the first eigenvalue, average between and within module correlation, and mean squared correlation for all traits (an overall measure of integration, see Porto et al. 2009). All these measures are related to the amount of integration in the system which in turn is inversely proportional to evolutionary flexibility (how close to the direction selection is pushing the population is able to respond or, in other words, the correlation between the selection gradient and evolutionary response, *sensu* Marroig et al. 2009). Accordingly, we can study the effects of these modularity/integration measures on selection reconstruction analyses.

In order to measure the impact of the extension approach over the correlation distribution in a sample matrix with respect to its "real" counterpart, we registered differences between estimated and "real" correlations, before and after applying the extension method for a single set of $A$-matrices generated from a sample of 50 individuals, thus illustrating what the extension method does to correlation estimates within each matrix.

### 3.3 *A study case* - Selection reconstruction in New World Monkeys skull evolution

To illustrate the application of the extension approach to evolutionary biology problems we carry through a selection reconstruction analysis on our New World Monkeys (NWMs) skull database. The data collection, measurements, sample sizes and procedures to obtain within-population phenotypic covariance matrices are described in Marroig and Cheverud (2001). All matrices used from now on were pooled for each node on the tree from within-population matrices observed in the terminal taxa (genus) using the phylogenetic tree of Platyrrhini (Wildman et al. 2009) as described in Marroig and Cheverud (2005, 2010). Average values for the 16 genera were used along with the phylogeny (Wildman et al. 2009) to obtain estimates of the direction of evolution ($\boldsymbol{\Delta\bar{z}}$) based on estimates of ancestral data (Marroig and Cheverud 2010). Using the Equation 1 we obtain the net selection gradients. Here $\boldsymbol{\Delta\bar{z}}$ is the difference vector between two nodes (or between a living genus and its ancestor) and we use the inverse of the phenotypic pooled within-group covariance matrix ($\boldsymbol{P^{-1}}$) as a proxy for $\boldsymbol{G^{-1}}$.

For this study case, our criterion for determining the cut-off point for the extension noise control were expanded in order to maintain the $\boldsymbol{\Delta\bar{z}}$ calculated using the noise-controlled $\boldsymbol{\beta}$ as close as possible to the original $\boldsymbol{\Delta\bar{z}}$ with respect to its magnitude and direction. Therefore, the cut-off point was chosen so that the percentile difference between the norms of both responses was not greater than 3%, and that their vector correlation stayed in the range of 0.97 to 1.00. After reconstructing selection gradients, we calculate their correlations with an isometric size vector in order to test whether these $\boldsymbol{\beta}$s represent selection in the direction of $\boldsymbol{P_{max}}$ or not. Also, by calculating the correlations between $\boldsymbol{P_{max}}$ and the response to selection, we can discern if this response was affected by the attractor effect of the direction of least genetic resistance or if the $\boldsymbol{\Delta\bar{z}}$ is a direct response to its corresponding $\boldsymbol{\beta}$. We also report the correlation between these reconstructed $\boldsymbol{\beta}$s and the corresponding $\boldsymbol{\Delta\bar{z}}$s (flexibility *sensu* Marroig et al., 2009), before and after controlling $\boldsymbol{P}$ for noise, in order to compare these flexibility values with those obtained by simulating 6,000 random $\boldsymbol{\beta}$'s, using Lande's equation to calculate $\boldsymbol{\Delta\bar{z}}$ and creating a distribution of flexibilities, akin to the approach taken by Marroig and Cheverud (2010).

## 4 Simulating G-matrices

We are also interested in the impact of noise control techniques in estimating $G$-matrices. Therefore, we would like to test if noise control techniques improve the estimation of $G$-matrices, comparing these estimates with "real" $G$-matrices before and after applying noise control techniques. Since "real" $G$-matrices are impossible to obtain from actual data, we address this question via a simulation-based approach.

Because $G$-matrices represent patterns of covariation between additive effects of genes over phenotypic traits, we are able to relate the features of these *loci* (like the frequencies and average effects of different alleles within each *locus*, for example) with the genetic variances and covariances of phenotypic traits (Falconer and Mackay 1996; Lynch and Walsh 1998; Kelly 2009). If these features are known, we can compare the real $G$-matrix with its estimate before and after applying noise control techniques with respect to reconstructing selection gradients.

We simulate three diploid populations, each comprised of 20,000 individuals. Each individual belonging to one of these populations is initially represented by a variable number of *loci* with purely additive effects (800 for the first population, and 1,150 for the second) affecting 40 traits. We assume initially that each locus has two possible alleles and for each individual we draw two alleles with equal probabilities; genotypic values for the three possible genotypes are -0.5 and 0.5 for both homozygotes, and zero for the heterozygote, as there is no dominance between each pair of alleles.

The 800 genes in the first population affect the 40 traits in a modular way, as they are divided into eight groups of 100 genes that affect groups of five traits. Each gene affects two traits within its given module; hence, the resulting $G$-matrix for this population has zero covariances between traits in different modules. The two other populations have genetic architectures that are modified from this basic framework. The second population has 350 additional genes that randomly affect two traits in each of the five modules simultaneously, so that its $G$-matrix had an average positive covariance between traits in different modules; therefore, this second $G$-matrix has a size factor (Bookstein et al. 1985) integrating all its modules. The third population has 800 genes controlling its traits, as the first, but on each module, two traits are affected by the same genes, so that their genetic correlation is one; hence, the resulting $G$-matrix has five dimensions with zero variance and does not have full rank.

After sampling all individuals belonging to both populations, we calculate average effects for each allele and breeding values for each individual (Falconer and Mackay 1996). In order to create phenotypes from these breeding values, we assign each trait to a score of heritability, randomly sampled from a F-distribution with 100 degrees of freedom for both numerator and denominator, centered to mean 0.4; we use this distribution here just to ensure that our random sample of heritabilities is non-negative and has a mean value compatible with those observed in morphological traits for endotherms (Mousseau and Roff 1987). Starting with these heritabilities, we calculate environmental variances for traits and use these variances to generate independent gaussian noise to the breeding values of all individuals. Therefore, environmental effects for all traits are uncorrelated, so that the $P$- and $G$-matrices for each population have strong similarities in their eigenstructure, as empirically found for morphological traits in mammals (see Cheverud 1995, 1996; Porto et al. 2009).

The individuals belonging to each population are paired randomly in 10,000 couples. These couples are used to produce a F1 generation, by sampling one of the two alleles for all *loci* independently, without linkage disequilibrium; genetic covariances between characters therefore arise only from pleiotropy. Each couple is used to produce two siblings, and the same procedure used to generate phenotypes in the parental generation was used in F1. Hence, each population comprised 10,000 families, each composed of sire, dam and two full siblings.

Since genealogical data for all populations are known, we are able to estimate $G$-matrices with REML methods (Shaw 1987, 1991; Lynch and Walsh 1998) by randomly sampling subsets of families within each population. Starting with a subset composed of 100 families, we increased sample size by 100 for each subsequent sample, until reaching 1,000 families. For each of these sample sizes, 100 *G*-matrices were calculated from resampling the total set of families.

After obtaining estimates for the *G*-matrix from phenotypic values, we are able to use these estimates to reconstruct selection gradients, applying 1,000 random $\boldsymbol{\beta}$s over the population *G*-matrix, and then using the $\boldsymbol{\Delta}\bar{\mathbf{z}}$s generated to reconstruct the original $\boldsymbol{\beta}$s with the estimated *G*-matrices. Therefore, we test the quality of the reconstructions with the same simulation scheme used on the theoretical matrices in the previous section (Figure 3). The shrinkage approach cannot be used in the REML estimated matrices because it demands actual measurements being correlated (in this case, estimated breeding values of each individual, which are not calculated in a simple REML estimation), so we restrict ourselves to the extension approach. All codes, matrices, software and simulated data can be obtained at http://dreyfus.ib.usp.br/gmarroig/extension/.

## 5  Results

A typical plot of observed eigenvalues of the *C*-matrix obtained for a population created from a uniform *C*-matrix is presented in Figure 1a, for different sample sizes. In the original matrix, eigenvalues are all equal. This sort of negative exponential eigenvalue plot is characteristic of morphological systems (Wagner 1984; Pavlicev et al. 2009a). In Figure 1b, notice the large 95% confidence interval and observed range in the estimates of the correlation (which should all be zero) in the original matrices compared to those where noise was controlled using the extended procedure. As expected by sampling theory, larger sample sizes are associated with smaller error (noise) in the estimates, but in all sample sizes the extended controlled estimates are substantially improved in comparison to the original ones.

Figure 4 shows the distribution of the correlations between true $\boldsymbol{\beta}$ values and those estimated directly or using the various noise control methods. We show results for a single type of matrix (C1, C2 and C3) and for different sample sizes, but the same pattern was found for other matrix types (see Supporting Information). Usually, both extension (VC) and shrinkage estimates of $\boldsymbol{\beta}$ are dramatically improved in regard to noise estimates.

Figure 5 shows how close to the best possible cut-off point (OC) noise control methods get in terms of the correlation between the estimated $\boldsymbol{\beta}$ and $\boldsymbol{\beta}_T$. Our VC is the one with the least difference to the OC (more points near zero, which indicate agreement with the OC cut-off point). The shrinkage method usually performs slightly worse, even though in a few samples this method is slightly better than the extended method using the OC in approaching $\boldsymbol{\beta}_T$. In practice both methods should be equivalent in their ability to improve our estimations of $C$-matrices. The original cut-off point of Twede and Hayden (2004) is inferior to other methods, estimating $\boldsymbol{\beta s}$ that are significantly different from the OC estimates in many cases. The original estimation without any noise control technique is also shown, and it can clearly be observed that this estimate is poor when compared to $\boldsymbol{\beta}_T$.

Figures 6 and 7 show how size variation (integration) and modularity affects the reconstructions. Within each matrix type (A, B, C and D) numbers 1, 2 and 3 indicate an increasing amount of variation associated with the first principal component (PC1). In general, matrices "1" do not have size variation with off-module correlations equal to zero while matrices 2 and 3 do have variation in size (see Supporting Information). Conversely, the number of modules is higher in matrices D (8 modules with 2 modules nested within each one to create a substructure – 8 x 2 = 16 modules), followed by matrices C (8 modules), B (4 modules) and A (only 2 modules). Therefore, within each matrix number (1, 2 and 3) matrices A > B > C > D have a decreasing amount of variation associated with their PC1 (Figure 6). The order of the matrices in terms of increasing integration is: D1 > C1 > B1 > A1 > D2 > C2 > D3 > B2 > A2 > C3 > B3 > A3 (see Figure 6). In our matrices, where the total amount of variation is similar, more variation in PC1 means that the last eigenvalues are even smaller, and thus even more difficult to estimate (Figure 2) which in turn affects our ability to obtain a reasonable inverse. Accordingly, matrices with high integration, low modularity and low flexibility (*sensu* Marroig et al. 2009) lead to estimates of $\boldsymbol{\beta}$ without noise control that are quite poor (see Table A1 on Supporting Information). Even controlling for noise there is an inverse relationship between magnitude of integration and efficiency of the selection reconstruction (Figure 6).

Regarding the effect of the extension method in estimating correlation values in each matrix, Figure 8 shows the distribution of deviations from population values for a selected set of three $A$-matrices. We can observe that the difference between population and estimated correlations increases in absolute value when we apply the extension method, especially in the $A3$-matrix, which has a more dominant size-factor (higher correlations). However, the same figure also shows that overall matrix structure does not change, so that applying the extension method does not affect the observation of modularity present in the population matrix. In fact, matrix structure is less patchy in those matrices controlled for noise since the general effect of noise control methods is to reduce the correlation values and thus control for spurious correlations.

## 5.1 G-Matrix

None of our REML $G$-matrix estimates for the first (size-free) population are positive-definite (which occurs when all eigenvalues of a given square matrix are strictly positive) even at our maximum sample size of 1,000 families. The occurrence of negative eigenvalues was common in our REML estimated matrices. For the third population, which is not positive-definite by definition, even some phenotypic matrices have negative eigenvalues. However, in the second population, affected by a size factor, 3% of our estimated $G$-matrices at sample size 700 are positive-definite; this percentage increases with larger samples so that, at our maximum sample size, 81% of these estimates are positive-definite.

Regarding selection reconstruction, raw estimates of $\boldsymbol{G}$ for both populations yielded poor estimates of $\boldsymbol{\beta}$ with respect to the correlation between those $\boldsymbol{\beta}$s and the ones applied over the population $G$-matrix. In the size-free population (Figure 9) these estimates are always poor, regardless of sample size; however, in the population with size effect, there is a slight improvement of reconstructions with respect to sample size (Figure 10). After applying either optimal (OC) or variance (VC) criteria, estimates of $\boldsymbol{G}$ for both populations improve the reconstructed $\boldsymbol{\beta}$'s, so that their correlation with the corresponding $\boldsymbol{\beta}_T$ have an asymptotic growth towards the maximum possible value of correlation with respect to sample size, as it can be observed in Figures 9 and 10, regarding the size-free and size-factor populations, respectively. Furthermore, the variance of the reconstructed $\boldsymbol{\beta}$ correlations is much lower in the noise controlled matrices (VC and OC) compared to the raw matrices without any form of noise control.

With respect to the phenotypic counterparts to estimates of $\boldsymbol{G}$, these P-matrices are much better behaved. Even without any noise control and at low sample sizes, *P*-matrices yield $\boldsymbol{\beta}$ reconstructions close to the true $\boldsymbol{\beta}$s considered. However, it appears that there is an upper limit to correlations between estimated and true $\boldsymbol{\beta}$s (around 0.9 for the size population and 0.75 for the size-free population). It is also noteworthy that, with respect to the size population, applying the extension method over phenotypic matrices slightly decreases the average correlation between real and estimated $\boldsymbol{\beta}$'s. The *P*-matrices in these simulations are so well estimated that the VC is unable to produce an improvement in $\boldsymbol{\beta}$ estimation, while the OC produces marginal improvement. This level of precision is of course highly unlikely in real studies, and the difference in $\boldsymbol{\beta}$ estimation is so slight that it would probably go unnoticed. Table A2 on the Supporting Information summarizes all results obtained for this section.

Results for the third population are presented in Supplemental Information (SI: Figure 2); however, it is noteworthy that such results are by no means different for those presented for the size-free population.

## 5.2 Study case

The noise and noise-controlled estimates of reconstructed $\boldsymbol{\beta}$'s for our study case are presented in Table A3 on the Supporting Information. Table I presents results regarding analyses comparing the directions of $\boldsymbol{\beta}$'s, $\boldsymbol{\Delta\bar{z}}$'s and $\boldsymbol{P_{max}}$. These results show that most responses to selection are significantly aligned with $\boldsymbol{P_{max}}$, in concordance with Marroig and Cheverud (2005). Also, in a dozen cases across the phylogeny of NWM, the direction of $\boldsymbol{\beta}$ after controlling for noise is also aligned with the direction of size. The bulk of such cases are concentrated on the Callithrinae clade (Figure 11).

Regarding estimates for flexibility before and after applying the VC (Table I), the original **β**'s (with noise) presents a range of values for flexibility between 0.11 and 0.45, usually outside the range of expected flexibilities for these matrices, obtained from the simulated distribution with known selection gradients; only three values of flexibility calculated with **β**'s without noise control along the phylogeny fall within the 95% confidence interval of the simulated distribution. When we use gradients reconstructed with noise control, the range of observed flexibility values becomes 0.45 to 0.76, which is compatible with the distribution of simulated flexibilities; all observed values along the phylogeny fall either within the 95 percentile empirical distribution of simulated values or slightly above.

## 6 Discussion

The difficulty in estimating eigenvalues of a uniform matrix, as shown in Figure 1, is a clear sign that noise control methods are a necessity when working with *C*-matrices. In fact, Figure 1 illustrates that with limited sample sizes, in a system where all traits are independent and equally variable, noise alone will lead to a negative exponential distribution of eigenvalues, a fact well-known (Anderson2003, Meyer and Kirkpatrick 2008). Furthermore, false positive correlation values with moderate to strong intensity tend to appear, an effect which is more pronounced in small sample sizes, as can be seen in Figure 1b.

Figures 4 and 5 illustrate the effect of noise and compares the shrinkage and extended methods, showing the influence of sample size on matrix estimation and consequently on selection reconstruction analyses. All methods perform better than *C*-matrix estimates without noise control by our criterion of how closely the reconstructed **β** is aligned to the true gradient. Although the effects of noise naturally diminish with increasing sample sizes, all noise control methods applied here perform better than the simple sample estimates. In fact, with sample sizes around 50 individuals, **β**'s reconstructed without controlling for noise are no more similar to the true selection gradient than expected by chance alone (see Table A1 in Supporting Information). After controlling for noise, however, selection gradient estimates are not only substantially improved but also present high and statistically significant similarity to the true gradients. Yet, it should be noted that the variance criterion (VC) developed here for the extended approach is in general the better performer as can be seen in Figure 5. The difference between the criteria suggested by Hayden and Twede (2002) and our VC can be observed in Figure 2. In general, the VC will select a smaller number of eigenvalues to be retained while the original (Hayden and Twede 2002) based on the minimum second derivative of the eigenvalues will usually identify the noise-floor at a later point over the ordered eigenvalues.

Matrices with enhanced modularity (larger number of modules and less integration) are less affected by noise and have both a higher similarity and a smaller variance in the correlations between the reconstructed $\boldsymbol{\beta}$s and true gradients (Figure 6). We can see the effect of modularity and integration in Figures 6 and 7, where increasing eigenvalue variance, either by raising between module correlations (thus increasing the first eigenvalue, matrices type 1, 2 and 3 respectively from lower to higher) or by having fewer relevant directions in the morphospace (with increasing degree of modularity, matrices type A, B, C and D) greatly affects the reconstruction analysis. In general, matrices that are less modular and have a strong integration (a substantial portion of the variability is concentrated on the first eigenvalue) will be more affected by noise than those matrices where dimensionality (number of modules) is higher and overall integration is lower (Figures 4, 6 and 7).

The connection between variational modularity and the difficulty in estimating selection gradients can be interpreted by noting that in less modular highly integrated systems a larger portion of the trait variation is shared with other traits and very little of the trait variation is not shared (common and trait-specific variation – see McGuigan and Blows 2010) making it harder to detect any signal in that specific part of the trait variation against noise. We show here that this type of noise can be readily identified with small eigenvalues of the $C$-matrix, and modularity aggravates this problem (Figures 2 and 6). This occurs because some directions on the morphological space contain less variation, due to the high correlation of traits within modules and consequently the concentration of the total variation in few dimensions. Thus, the last eigenvalues are correspondingly smaller and carry only a tiny fraction of the total variation in the system (Pavlicev et al., 2009a). Since smallest quantities had relatively more error in its estimates than larger quantities (just because they are small and approach the error inherent to both observer and measuring device) it is easy to see why the last eigenvalues have a low signal-to-noise ratio. Just to give an idea of the noise control effect on the millimeter scale of real measures taken from a sample of marmosets skulls we compared the observed values in a sample of 502 skulls with the values after controlling for the noise in the C-matrix (adjusting eigenvalues via extended approach and re-projecting the data back to the original basis). The average difference in mm between observed data and noise-controlled data is around 0.12 mm which is well within observer plus measuring device (0.05 mm accuracy of the 3D digitizer) error.

When we consider the effect of extension on the individual correlation estimates (Figure 8), it is clear that by controlling noise we are increasing to some extent the variation in the system (the small eigenvalues are increased), and so estimated correlation values tend to become smaller than true values, an effect more pronounced in size-dominated matrices. It is noteworthy, however, that this bias represents a small one when compared to the absolute value of correlations in these matrices, within and between modules. In the example provided in Figure 8, intra-modular correlations in the estimated $A3$-matrix reduce to an average of 0.6 when the extension method is applied. This is a small change when compared to the real value of 0.8, and the reduced value is still statistically different from zero at a sample size of 50 individuals. At same time, small spurious correlations that reflect noise in the system are controlled by extension and consequently the structure of the matrix is a much better representation of the true pattern structure than the raw matrix (see Figure 8, especially the left panel, matrix $A1$ where off-modules correlation should be zero). This illustrates the effects of the extension approach over the eigenvalue distribution and its impacts on $C$-matrix structure. As we extend the lower eigenvalues, we reduce the variance of their distribution, effectively controlling for its bias, while introducing a small bias in the eigenvalue mean. Our argument here is that this bias in eigenvalue mean does not affect $C$-matrix structure as much as a bias in eigenvalue variance, as shown above.

Regarding results with simulated $G$-matrices for the three populations, it is clear that the extension approach improves dramatically estimation of selection gradients, regardless of sample size. Often, estimated $G$-matrices will have negative eigenvalues, representing linear combinations of traits that have negative additive genetic variances (Hill and Thompson 1978); hence, these matrices cannot be considered to be true

*C*-matrices. These negative eigenvalues appear in estimated *G*-matrices due to impossible values of estimated pairwise correlations, either because they fall outside the range of -1 to 1, or because there are combinations of partial correlations for certain subsets of traits that are impossible in terms of variance partitioning. Therefore, *G*-matrices estimated through separate REML models for each of its variances and covariances need to be corrected for bias in its eigenvalues; our argument here is that the extension method is the best correction because it preserves modularity structure in any *C*-matrix estimate, be it phenotypic or genetic; the extension method is also a simple method in terms of implementation when compared to the shrinkage method.

In our simulated population, virtually all *G*-matrix estimates in the size-free population have negative eigenvalues where there are positive eigenvalues in the population *G*-matrix. Since the eigenvalue distribution of a *G*-matrix estimate is used for estimating the dimensionality of the population *G*-matrix (e.g.: Mezey and Houle 2005; Hine and Blows 2006; Pavlicev et al. 2009b), we suggest that one should proceed with caution when interpreting negative eigenvalues as dimensions without genetic variation (see also Pavlicev et al. 2009b).

In the size-factor population, some of the *G*-matrix estimates at higher sample sizes become free of negative eigenvalues, which makes these estimates more reliable even without any correction for their eigenvalue distribution. This is by no means a property of biological systems that have size variation; it is only a byproduct of our simulation design. Since heritabilities are fixed, genetic variation in size increases the overall genetic and phenotypic variances in the system, reducing environmental variances. Therefore, estimates for genetic variances and covariances are easier to obtain in the size-factor population, simply because they are higher than those in the size-free population (Young et al. 2010).

A corollary of our simulations is that, when environmental effects are low and normally-distributed and interactions between genetic and environmental effects are nonexistent, as it is the case here, *P*-matrices produce better estimates of selection gradients than *G*-matrices without using any noise control techniques, as can be seen in the leftmost panels of Figures 9 and 10. This is consistent with Cheverud's conjecture (Cheverud 1988; Roff 1995; Reusch and Blanckenhorn 1998; Waitt and Levin 1998; Porto et al. 2009;

Dochtermann 2011), which suggests that *P*-matrices are good approximations of their genetic counterparts for evolutionary analyses, given that environmental effects are low or have similar structure to *G*. This result reflects the fact that, under the conditions of our simulation, *P*-matrices are estimated with less noise due to their large sample sizes than their corresponding *G*-matrices due to the familial structure (Roff 1995).

In all the examples based on simulations, sample sizes considered as adequate for studies based on museum or laboratory collections ($n = 100$ or $n = 200$) still present a bias, with the first eigenvalues being overestimated and the last ones underestimated. Even when working with matrices with moderate rank, only huge, realistically unobtainable samples lead to ML estimates with reasonable error in eigenvalues. This is an important message for researchers about an effect which can be exacerbated in situations where the number of estimated parameters is larger than or close to sampled specimens, as is common in geometric morphometrics nowadays. The situation is even more dramatic in gene co-expression studies where thousands of dimensions (genes) might be correlated with only a few individuals. This has already been noticed in other fields in which the error in estimation is more evident, such as genomics or even economics, where covariance matrices that are not positive definite are obtained if data is used directly, and thus easily perceived as in need of improvement (Ledoit and Wolf 2003; Schafer and Strimmer 2005; see also Efron 1982 for a discussion on the limitations of ML estimators). While the problem has been noticed, noise control is not yet a common practice in genomic studies (e.g.: Oldham et al. 2006; Nowick et al. 2009). Both shrinkage and extended approaches can be used in estimates of variance/covariance or correlation as a way of controlling noise and improving representations of population patterns. We illustrate this in the context of selection reconstruction analyses exploring both modularity and sampling in signal-to-noise ratio in simulated datasets as well as estimating selection gradients in New World Monkeys skull evolution. We shall now consider these results.

Previously, Marroig and Cheverud (2004, 2005) showed that skull evolution of New World Monkeys was predominantly under natural selection and that the adaptive radiation of this group was dominated by size changes (and allometrically correlated shape changes) associated with dietary differences. However, net selection gradients reconstructed were not similar to size, suggesting that size evolution was a by-product of the attractor effect of the line of least resistance and its long term conservation (Marroig and Cheverud 2005). Yet, net selection gradients reconstructed previously (Figure 4 in Marroig and Cheverud 2005 and supplementary Table 3) present vector correlations (between 0 and 0.44), with the observed evolutionary responses ($\boldsymbol{\Delta \bar{z}_{observed}}$) outside the range normally observed in our simulation (0.4 to 0.85) of evolutionary responses produced by random selection vectors (Table 1, see Marroig and Cheverud 2010, Marroig et al. 2009), suggesting that they were poor estimates of the true net-$\boldsymbol{\beta}$ due to noise in those C-matrices. Furthermore, Marroig and Cheverud (2010) developed a method based on random selection simulations that allows to test whether a particular observed case of evolutionary change along a size dimension can be explained by the line of least resistance attractor effect. In the New World monkeys skull evolution study they show that at least in a third of the total cases the observed size/allometry evolutionary change probably result from selection to some extent aligned with size. We confirm this finding here by removing noise from matrices estimates prior to reconstructing net selection gradients (Table 1). Five of the six cases identified by Marroig and Cheverud (2010) as being under size selection (*Callicebus*, *Cebus*, *Saguinus*, Callithrix and *Cebuella*) also present here a significant correlation of the net-$\boldsymbol{\beta}$ with a size vector (Table 1, Figure 11). Furthermore, there is a clear signal of size selection in branches corresponding to marmosets and tamarins where a long term trend for reducing overall body size is observed (Figure 11).

**Conclusion**

It is important that we realize that the consequences of not controlling noise from our estimated matrices would not only impact selection reconstruction but any statistical inference based on matrix estimates and, particularly, in a very pronounced way, inverted matrices. Thus, while we illustrate the problem here within the context of selection reconstruction analysis, the problem (noise) and solutions to control it (extension and shrinkage) have far broader implications. Both traditional Euclidean and Geometric morphometrics make extensive use of covariance matrices and invert them routinely. In ecology, genomics and genetics many procedures involve matrix inversion. In these contexts, using ML estimates lead to results that may be extremely different from reality. For example, in phylogeography and phylogenetic analyses we can think about historical relationships among units (whether populations or species) in terms of a covariance matrix of ancestry (Cavalli-Sforza and Piazza 1975). These matrices will be estimated with some degree of noise and the problem so far has been largely ignored. Noise control methods discussed here can be useful in these applications and we plan to explore that in a subsequent contribution. In evolutionary ecology, methods dealing with demography and population growth are usually represented in form of covariance or correlation matrices (Caswell, 2001, 2008) and the same problem will be present in such estimates. In fact, given the central importance of correlation and variance/covariance matrices in quantitative biology we suggest that researchers should incorporate noise control techniques in matrix estimation as a routine procedure and to fully incorporate the extension approach in the housekeeping statistical toolkit.

## 7 Figure Captions

Figure 1: Bias introduced in *C*-matrix estimates by sample size from a population originated with a multivariate normal distribution with uniform covariance matrix: (a) eigenvalue distribution for a series of matrices estimated for a series of increasing sample sizes; (b) distributions of correlations observed in sample *C*-matrices estimated with increasing sample size.

Figure 2: Eigenvalues for a C3 class matrix (See Section Simulation and Bootstrap). Sample eigenvalue estimation mean and coefficient of variation taken from 1000 bootstrap matrices with 50 individuals taken from a multivariate normal distribution with C3 as the covariance matrix.

Figure 3: Simulation scheme for testing noise control techniques in selection reconstruction analysis.

Figure 4: Correlation between true and estimated $\boldsymbol{\beta}$'s for the matrices type C1, C2 and C3. See Supporting Information and text for detailed description of the matrices.

Figure 5: Comparison of each method with the OC. More points near zero indicate complete agreement in the correlation with using the OC and each method.

Figure 6: Correlation between true and estimated $\boldsymbol{\beta}$'s without controlling for noise with N=500 as a function of total integration

Figure 7: Correlation between true and estimated $\boldsymbol{\beta}$'s using the OC. Here we see how having a first principal component (size) accounting for most of the variability in the sample affects the reconstruction analysis. The larger the first PC, the harder it is to obtain a reasonable reconstruction of the selection gradient.

Figure 8: The effect of the extension method over correlation values. In the first line of graphics, we show the distribution of correlation values for A1, A2, and A3-matrices before and after applying the extension method, with correlations between and within each module set apart. Lines on each graph indicate the true correlation values for each population; dotted lines indicate extramodular correlations and dashed lines indicate the intramodular ones. In the three subsequent lines, we show graphic representations of these matrices: the population matrix, the estimated matrix without noise control, and the estimated matrix after applying the extension method. The grayscale used in these matrices represent the maximum correlation level as white and the minimum as black, with intermediate values as shades of grey.

Figure 9: Reconstruction of selection gradients using the estimated G- and P-matrices for the size-free population, with and without noise controlling methods. Correlations between true and estimated $\boldsymbol{\beta}$'s using both optimal (OC) and variance (VC) criteria.

Figure 10: Reconstruction of selection gradients using the estimated G- and P-matrices for the size-factor population, with and without noise controlling methods. Correlations between true and estimated $\boldsymbol{\beta}$'s using both optimal (OC) and variance (VC) criteria.

Figure 11: Association between reconstructed $\boldsymbol{\beta}$'s and the direction of size across the phylogeny of Platyrrhini monkeys. The strength of correlation between these vectors is indicated by the color scale and by circles and squares, whether such correlation is negative or positive, respectively.

## 8 Table Caption

Table 1: Reconstruction of noise-controlled $\boldsymbol{\beta}$'s across the Platyrrhini phylogeny and their relationship with $\boldsymbol{\Delta\bar{z}}$'s and $\boldsymbol{P_{max}}$.

## 9  Supporting Information Captions

Table A1: Table with all means and standard deviations for bootstrap simulations.

Table A2: Table with all means and standard deviations for *G*-matrix simulations.

Table A3: Selection gradients with noise and controlled for noise in the P-matrices along with the $\boldsymbol{\Delta\bar{z}}$'s. Node numbers correspond to figure 11. Also, selection gradients reconstructed after controlling for noise in the Saguinus G-matrix are presented.

Table A4 - All 14 population matrices from which samples were draw from a multivariate normal distribution.

Figure A1: Graphical representation of the matrices used in section "Simulation and Bootstrap".

Figure A2: Reconstruction of selection gradients using the estimated G- and P-matrices for the reduced-rank population, with and without noise controlling methods. Correlations between true and estimated $\boldsymbol{\beta}$'s using both optimal (OC) and variance (VC) criteria.

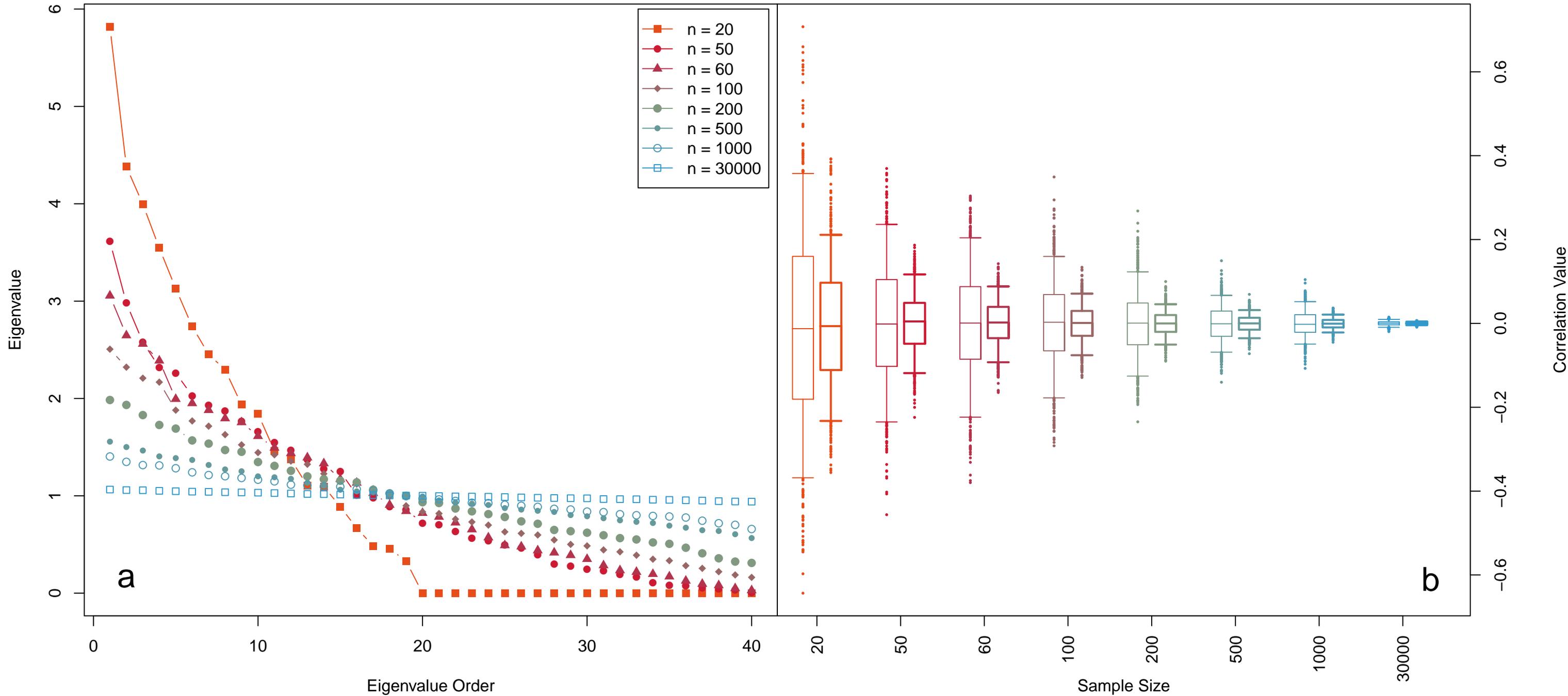

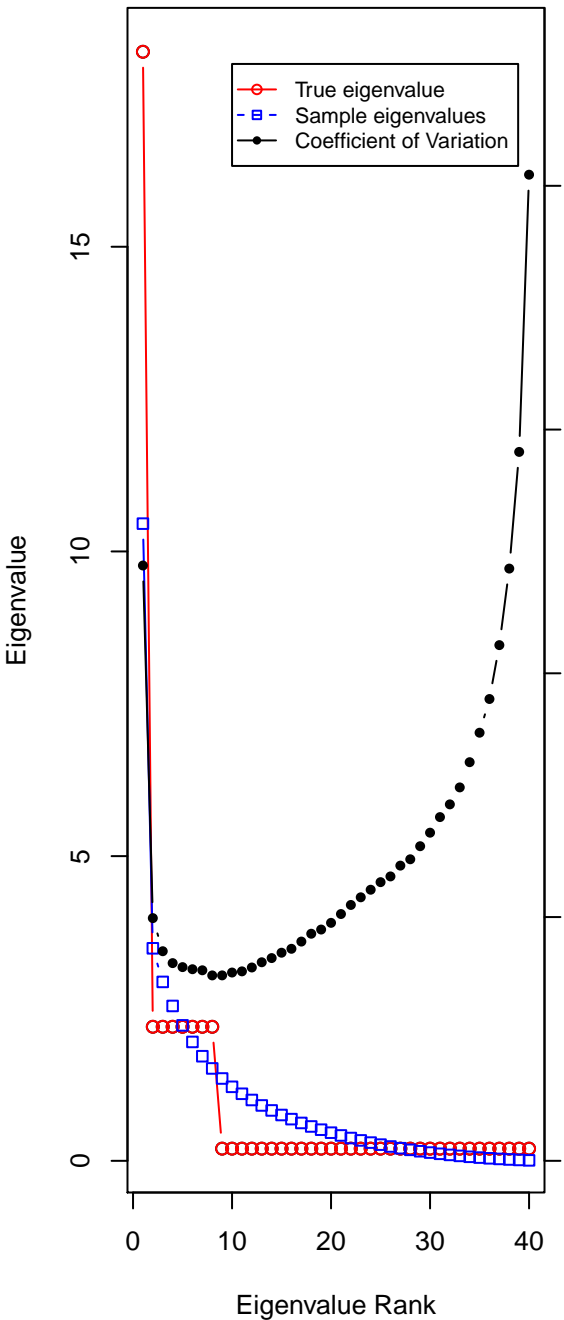

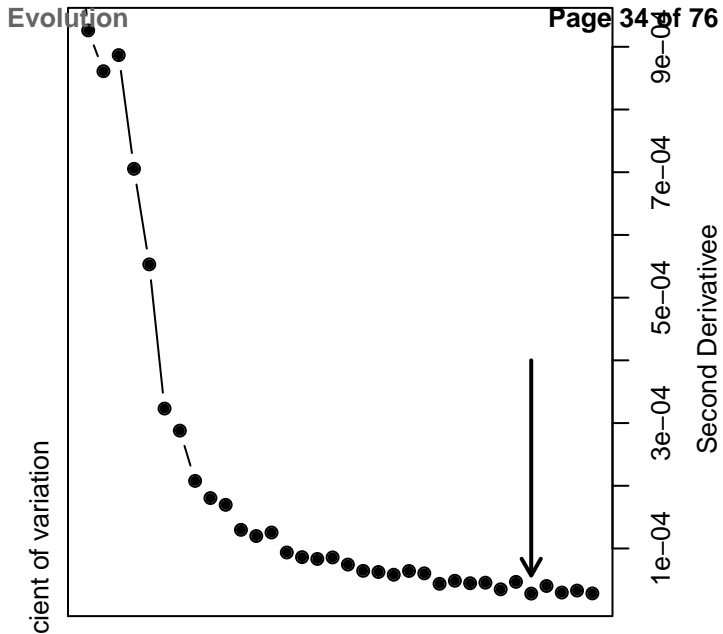

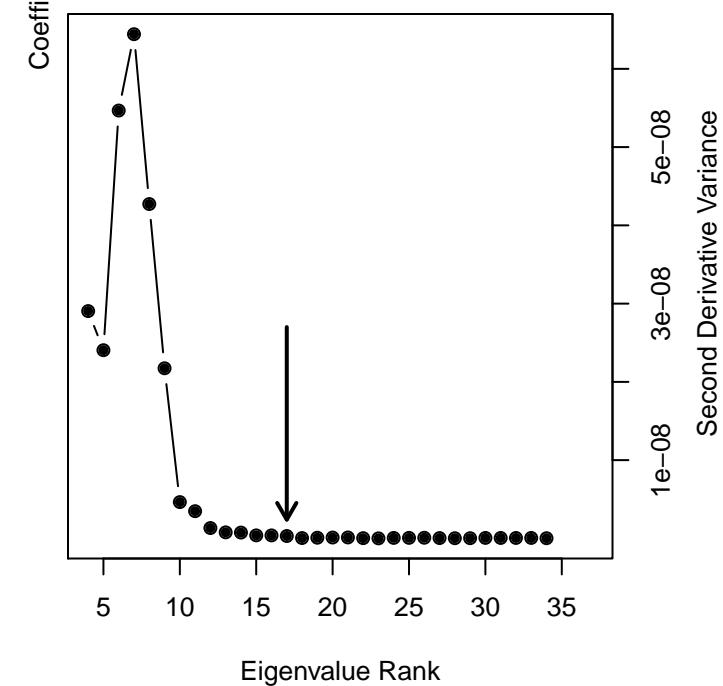

$$\text{Simulated } \beta_T$$

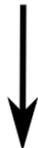

$$\Delta z = C_T \beta_T$$

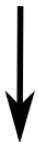

$$\beta_s = C^{-1}_{sample} \Delta z$$

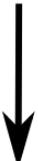

$$Corr(\beta_s, \beta_T)$$

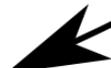

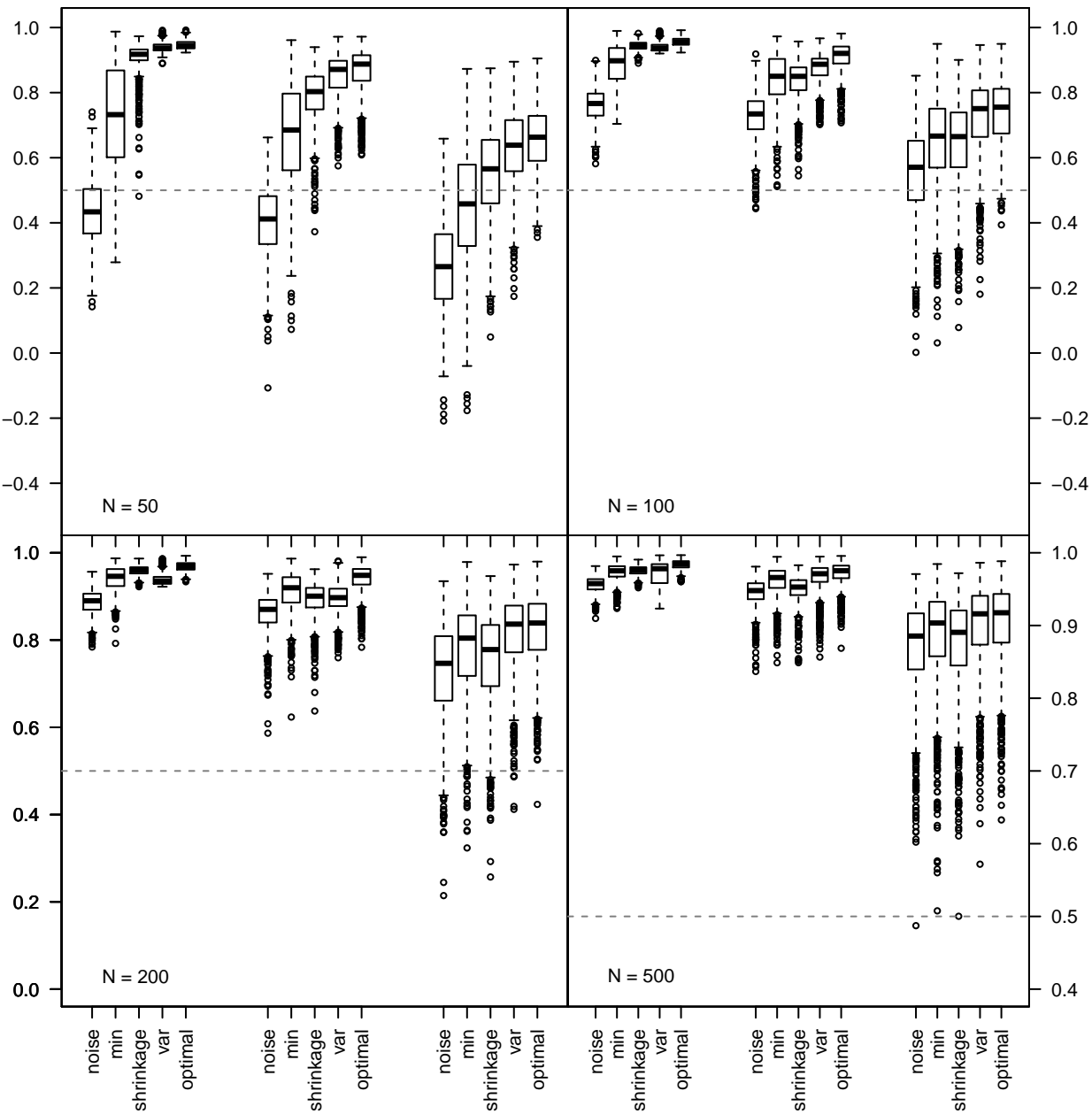

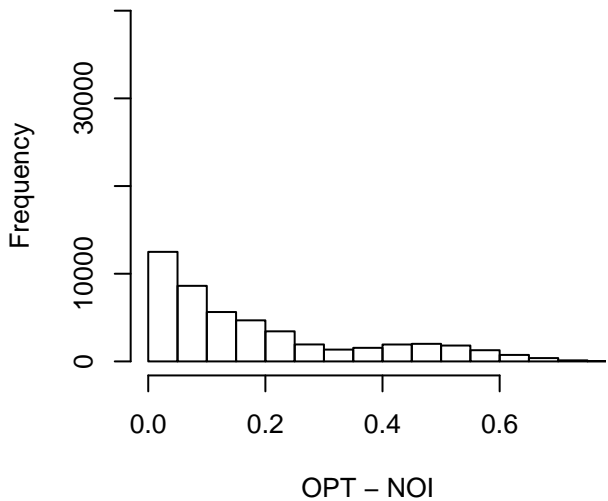

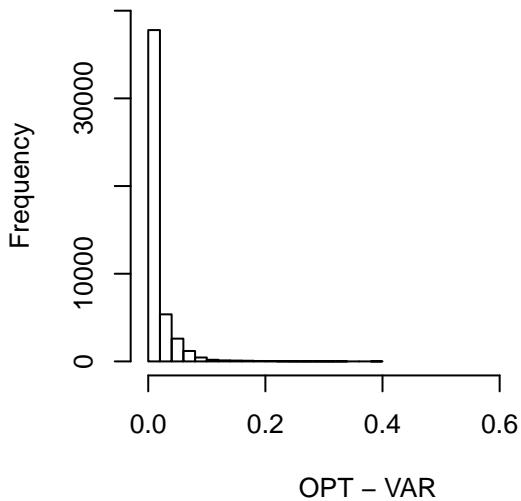

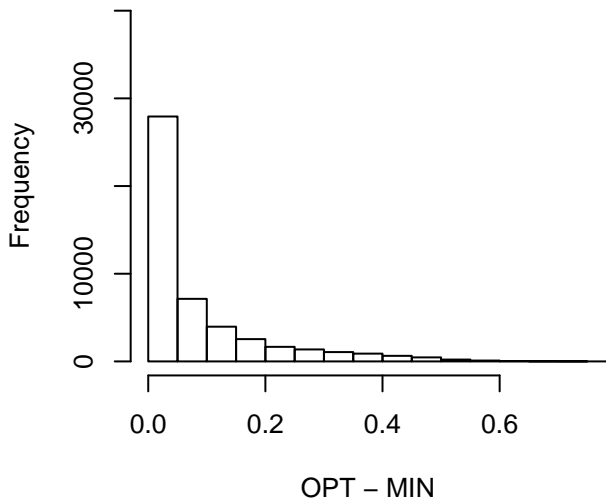

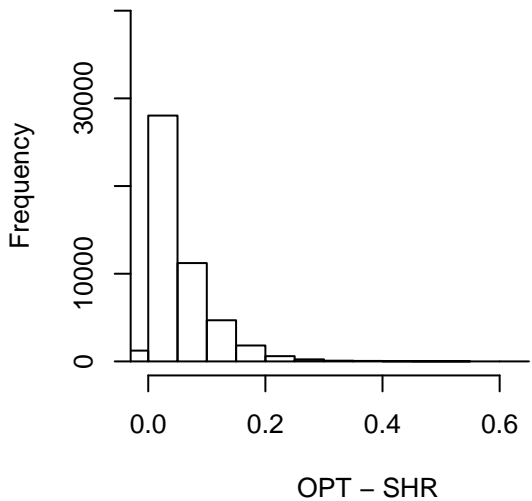

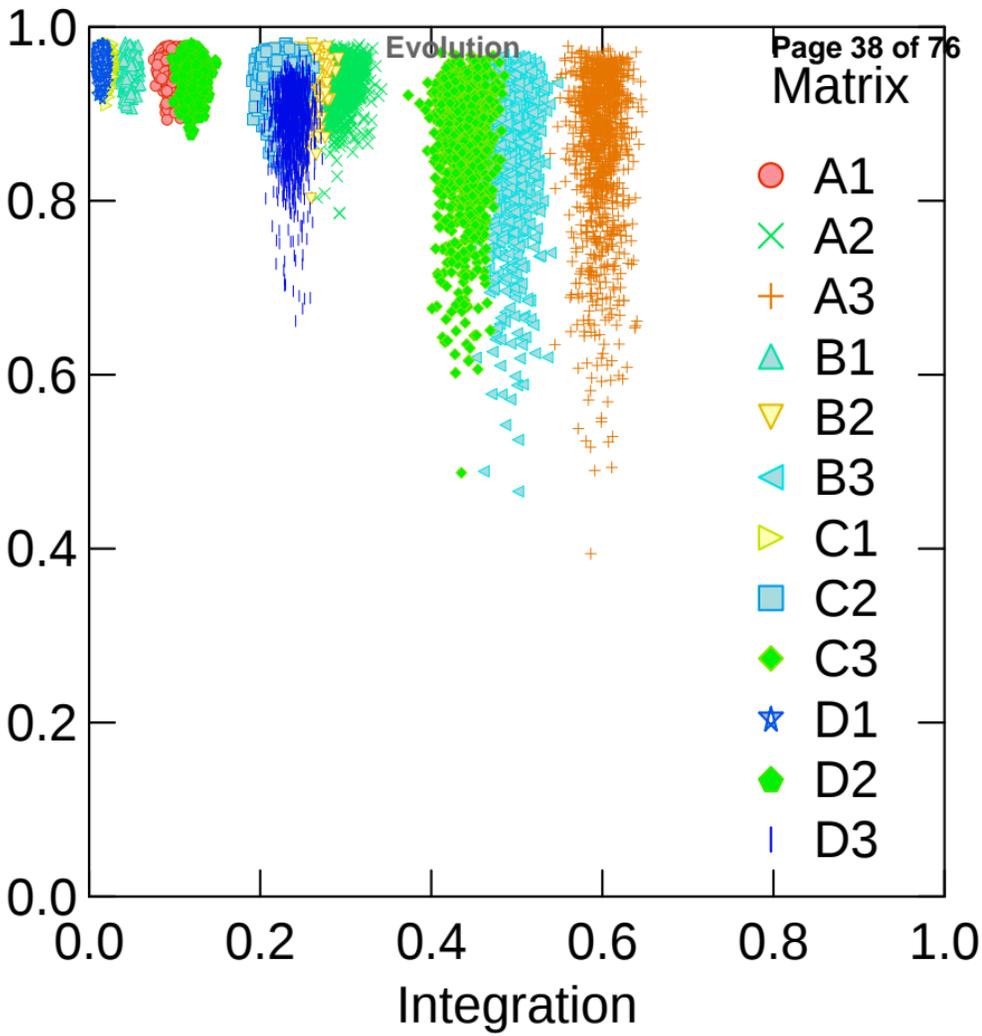

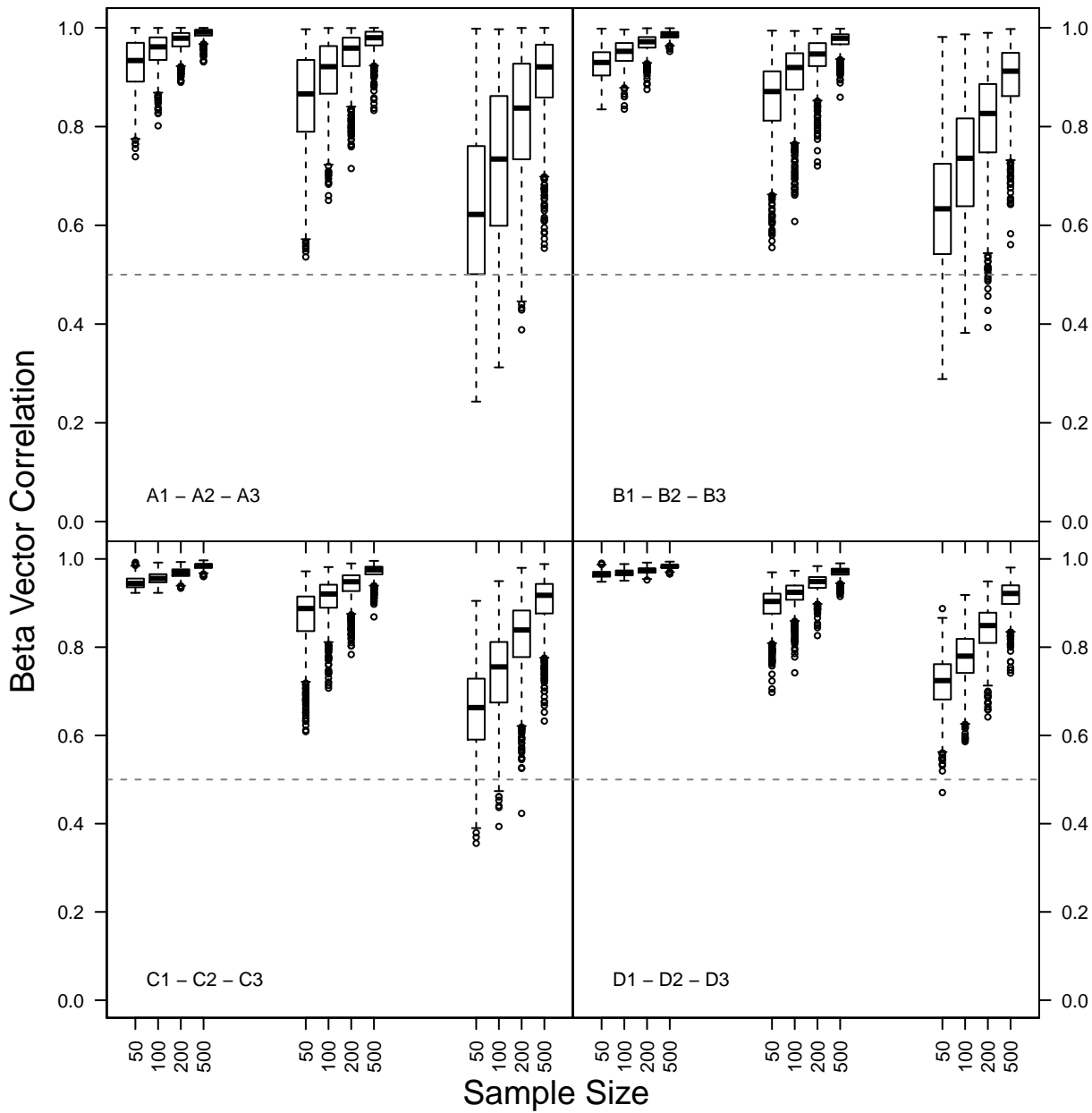

## A1

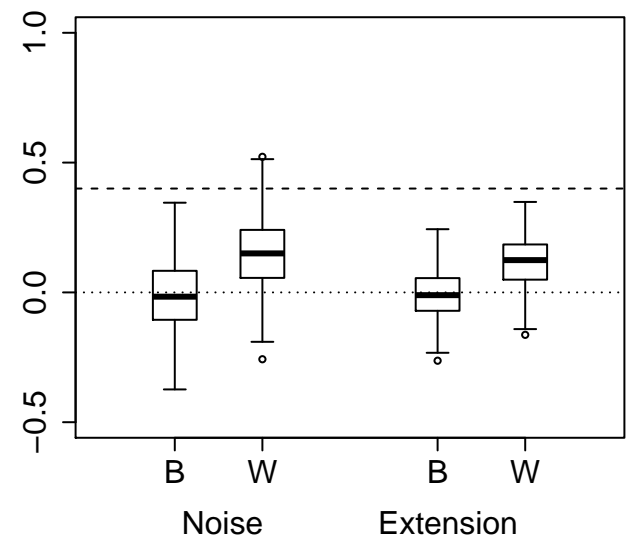

## Evolution

## A2

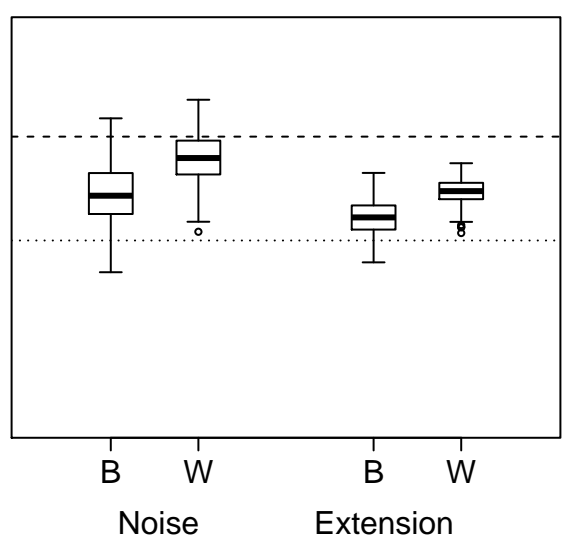

## A3

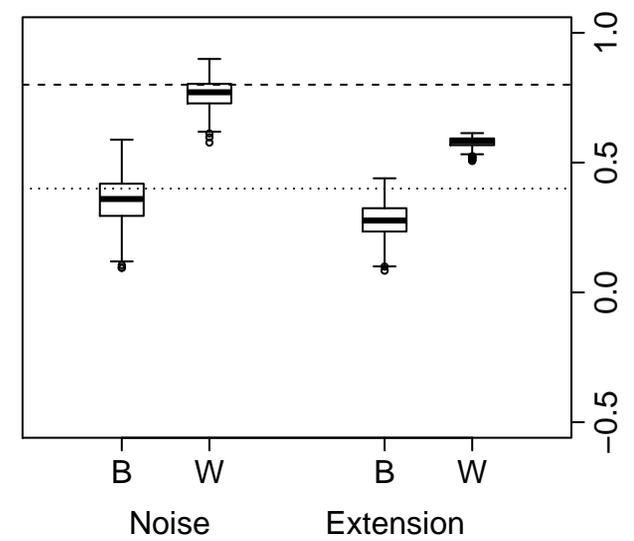

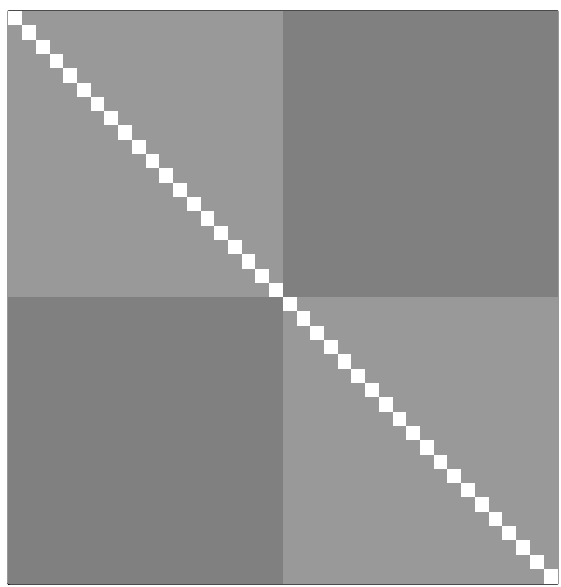
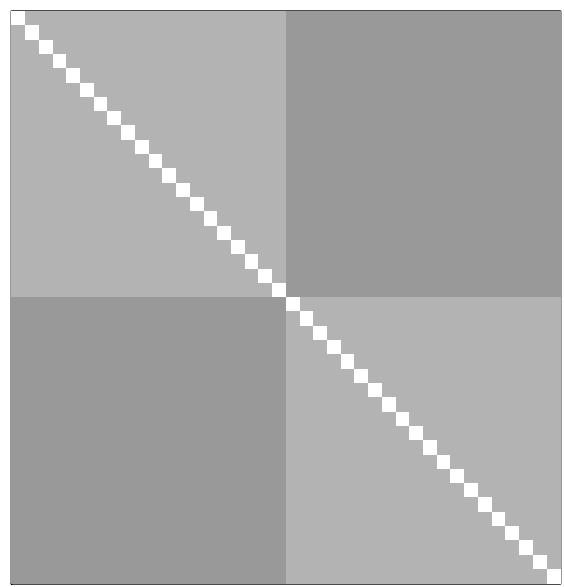
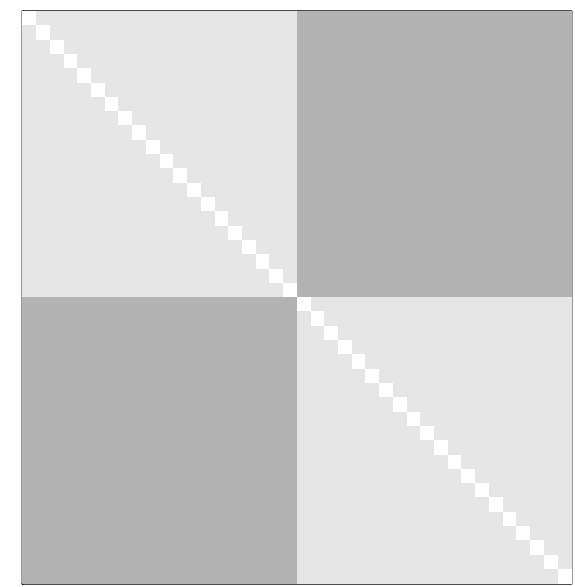

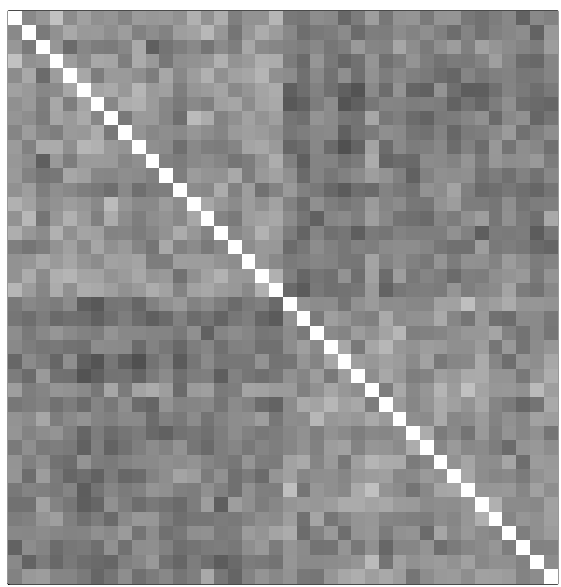
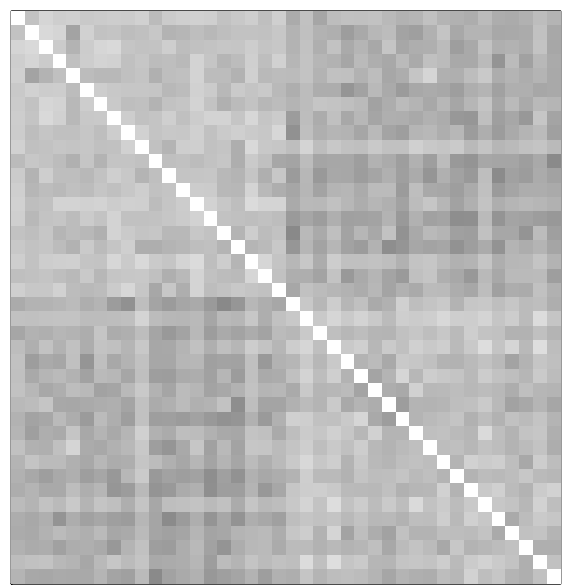
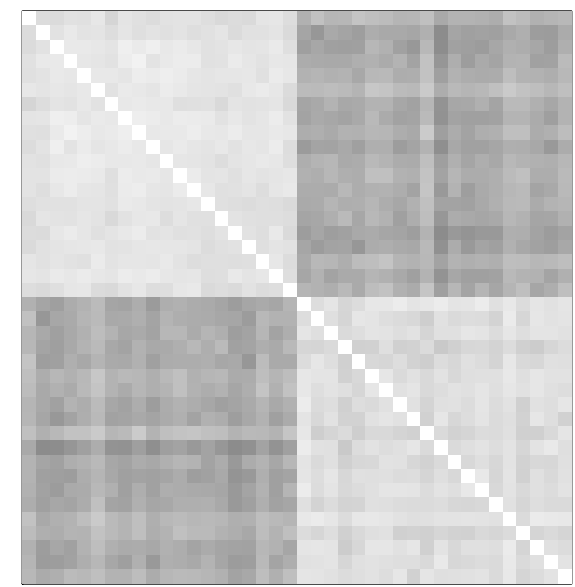

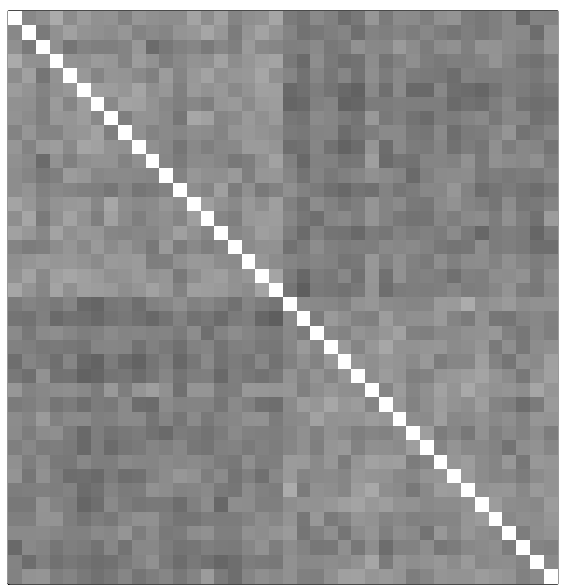
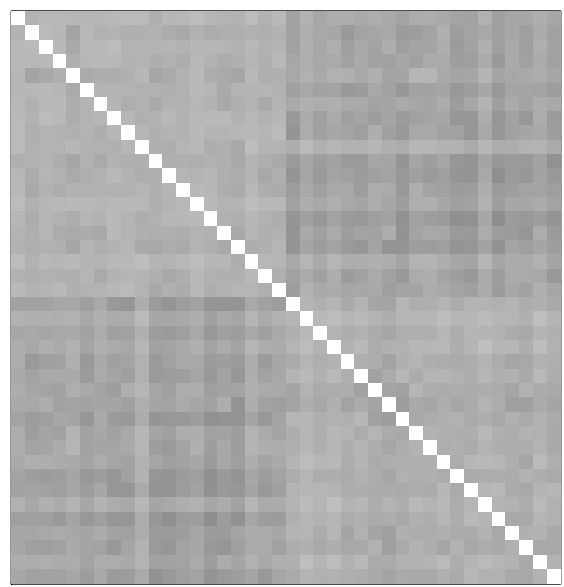
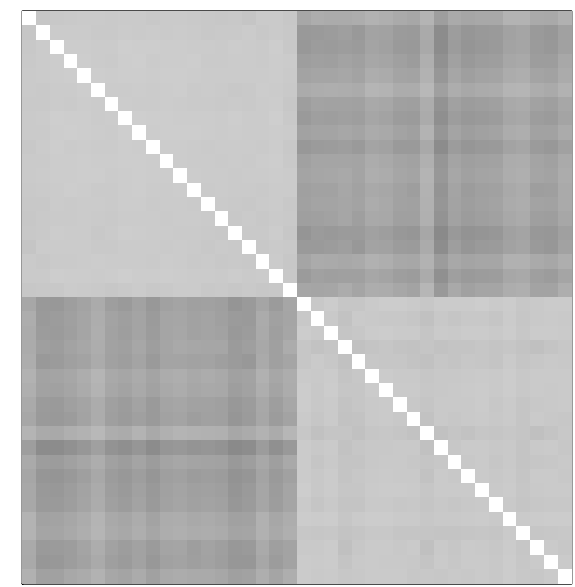

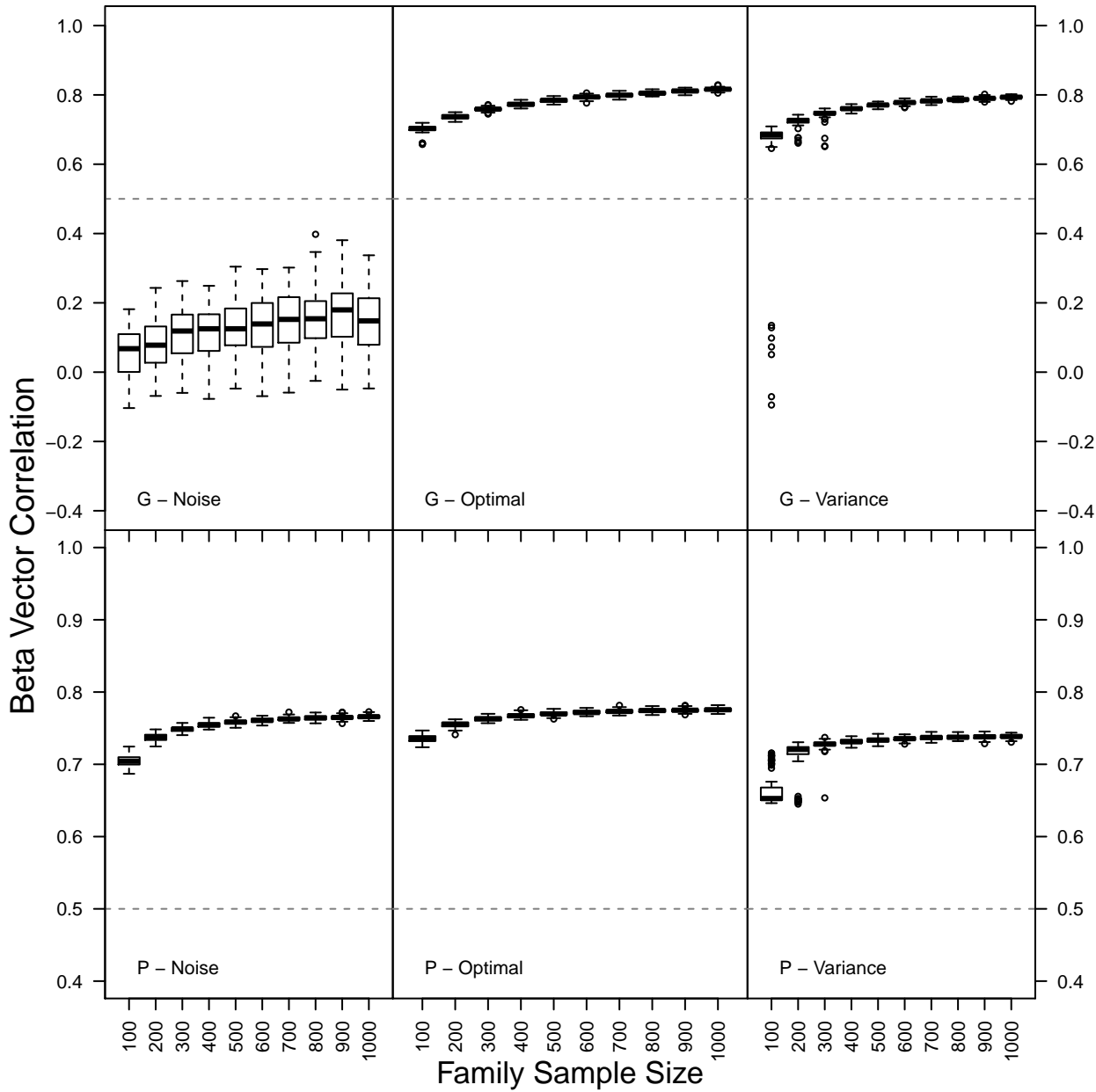

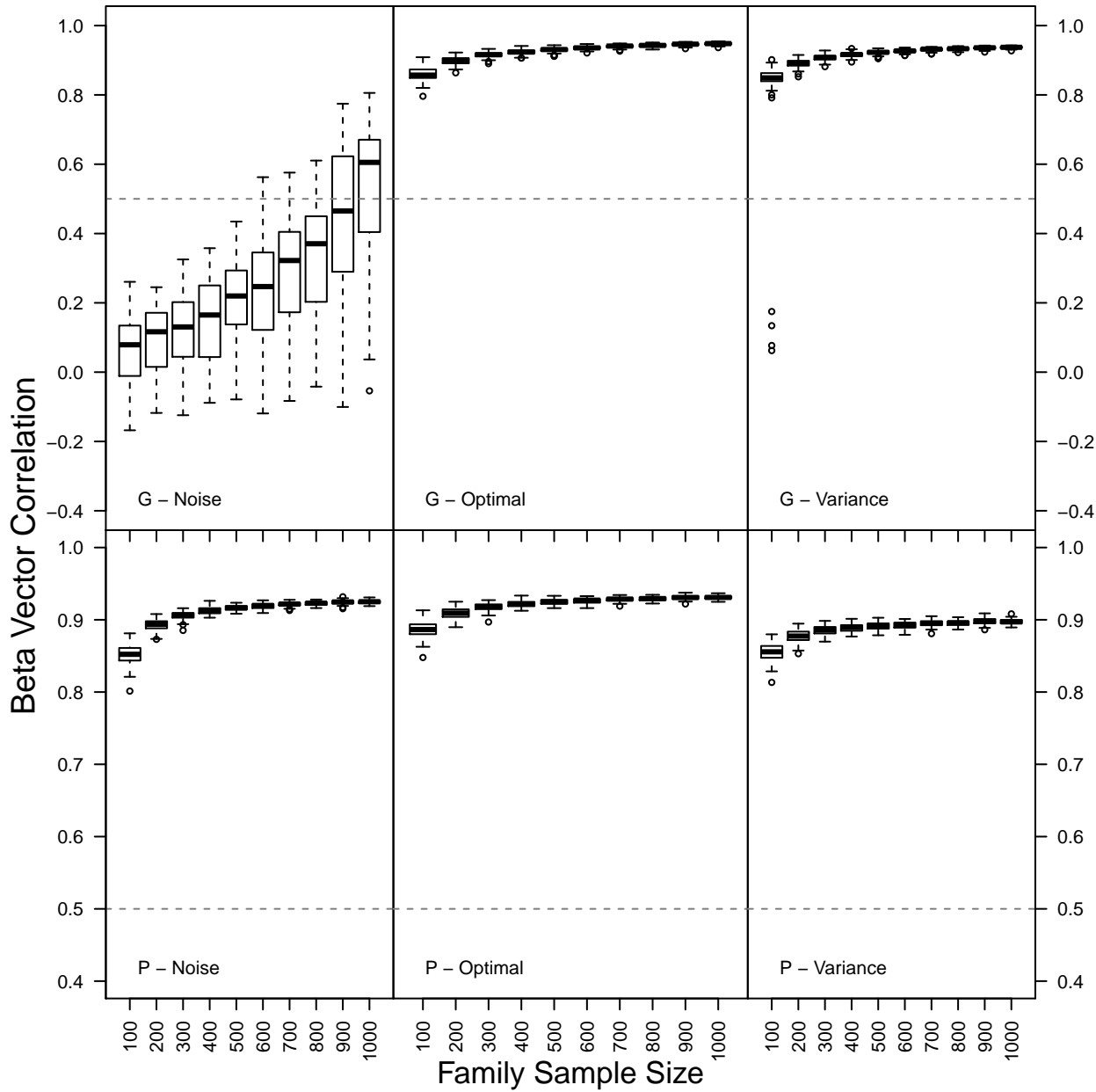

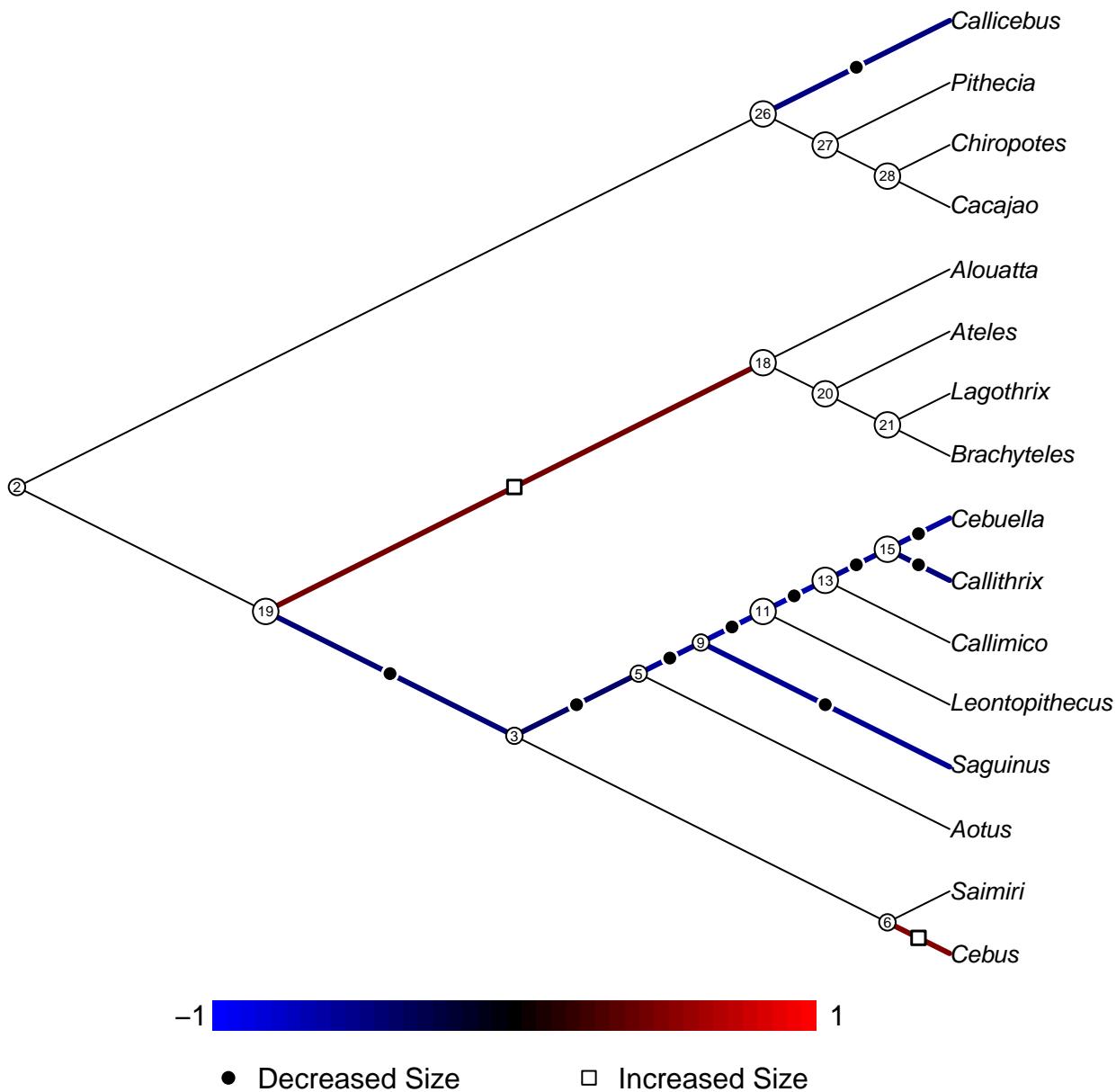

Planilha1

| Node | PC's Retained | % dif. | $\Delta z$ x $\Delta z$_observed | $\beta$ x Size | Constraints |
|---|---|---|---|---|---|
| 9-5 | 7 | 1.554 | 0.993 | **-0.589** | **-0.901** |
| 15-13 | 12 | 0.808 | 0.997 | **-0.594** | **-0.934** |
| 13-11 | 8 | 1.266 | 0.994 | **-0.701** | **-0.964** |
| 11-9 | 8 | 1.533 | 0.993 | **-0.690** | **-0.921** |
| 6-3 | 10 | 1.580 | 0.995 | 0.125 | **0.729** |
| 5-3 | 10 | 2.845 | 0.988 | **-0.409** | **-0.865** |
| 19-2 | 24 | 2.990 | 0.992 | 0.110 | **0.548** |
| 19-3 | 9 | 1.162 | 0.995 | **-0.474** | **-0.927** |
| 18-19 | 9 | 1.429 | 0.993 | **0.453** | **0.904** |
| 20-18 | 10 | 1.768 | 0.992 | 0.375 | **0.739** |
| 21-20 | 18 | 2.522 | 0.991 | 0.180 | **0.795** |
| 26-2 | 24 | 2.990 | 0.992 | -0.110 | **-0.557** |
| 28-27 | 18 | 1.213 | 0.994 | 0.258 | **0.698** |
| 27-26 | 13 | 2.895 | 0.984 | 0.275 | **0.787** |
| Pithecia-27 | 18 | 2.771 | 0.990 | -0.212 | -0.284 |
| Callicebus-26 | 8 | 2.269 | 0.991 | **-0.512** | **-0.843** |
| Ateles-20 | 17 | 2.865 | 0.990 | 0.200 | **0.640** |
| Alouatta-18 | 16 | 1.530 | 0.995 | -0.046 | **0.727** |
| Aotus-5 | 15 | 2.563 | 0.991 | -0.071 | **-0.736** |
| Saguinus-9 | 8 | 3.000 | 0.987 | **-0.572** | **-0.868** |
| Leontopithecus-11 | 21 | 2.485 | 0.991 | -0.117 | **0.719** |
| Callimico-13 | 24 | 2.969 | 0.988 | 0.032 | -0.478 |
| Callithrix-15 | 12 | 1.293 | 0.995 | **-0.479** | **-0.886** |
| Cebuella-15 | 12 | 0.783 | 0.997 | **-0.623** | **-0.925** |
| Cebus-6 | 5 | 2.077 | 0.990 | **0.507** | **0.864** |
| Saimiri-6 | 12 | 1.708 | 0.994 | -0.197 | **-0.844** |
| Brachyteles-21 | 18 | 2.866 | 0.988 | 0.213 | **0.744** |
| Lagothrix-21 | 18 | 2.774 | 0.988 | -0.053 | 0.263 |
| Cacajao-28 | 13 | 2.504 | 0.989 | 0.333 | **0.746** |
| Chiropotes-28 | 24 | 2.961 | 0.988 | -0.072 | -0.017 |

Planilha1

Flexibility

| Noise | Control | Simulation | | | |
|---|---|---|---|---|---|
| | | Lower Bound | Average | Upper Bound | SD |
| 0.257 | 0.718 | 0.374 | 0.511 | 0.665 | 0.075 |
| 0.220 | 0.634 | 0.416 | 0.547 | 0.684 | 0.069 |
| 0.283 | 0.760 | 0.415 | 0.548 | 0.683 | 0.070 |
| 0.347 | 0.764 | 0.439 | 0.590 | 0.732 | 0.078 |
| 0.249 | 0.604 | 0.374 | 0.511 | 0.665 | 0.075 |
| 0.279 | 0.593 | 0.374 | 0.511 | 0.665 | 0.075 |
| 0.448 | 0.547 | 0.372 | 0.527 | 0.691 | 0.084 |
| 0.278 | 0.613 | 0.366 | 0.520 | 0.685 | 0.082 |
| 0.273 | 0.612 | 0.366 | 0.520 | 0.685 | 0.082 |
| 0.320 | 0.708 | 0.343 | 0.505 | 0.685 | 0.091 |
| 0.296 | 0.501 | 0.387 | 0.512 | 0.659 | 0.070 |
| 0.448 | 0.547 | 0.372 | 0.527 | 0.691 | 0.084 |
| 0.281 | 0.627 | 0.383 | 0.546 | 0.703 | 0.086 |
| 0.315 | 0.618 | 0.384 | 0.549 | 0.710 | 0.088 |
| 0.258 | 0.759 | 0.383 | 0.546 | 0.703 | 0.086 |
| 0.298 | 0.703 | 0.384 | 0.549 | 0.710 | 0.088 |
| 0.363 | 0.659 | 0.387 | 0.512 | 0.659 | 0.070 |
| 0.370 | 0.592 | 0.343 | 0.505 | 0.685 | 0.091 |
| 0.274 | 0.580 | 0.374 | 0.511 | 0.665 | 0.075 |
| 0.331 | 0.750 | 0.439 | 0.590 | 0.732 | 0.078 |
| 0.426 | 0.607 | 0.415 | 0.548 | 0.683 | 0.070 |
| 0.327 | 0.566 | 0.416 | 0.547 | 0.684 | 0.069 |
| 0.225 | 0.661 | 0.422 | 0.553 | 0.690 | 0.069 |
| 0.245 | 0.469 | 0.422 | 0.553 | 0.690 | 0.069 |
| 0.110 | 0.738 | 0.302 | 0.427 | 0.577 | 0.071 |
| 0.203 | 0.480 | 0.302 | 0.427 | 0.577 | 0.071 |
| 0.219 | 0.570 | 0.405 | 0.548 | 0.702 | 0.077 |
| 0.226 | 0.680 | 0.405 | 0.548 | 0.702 | 0.077 |
| 0.276 | 0.643 | 0.386 | 0.542 | 0.695 | 0.081 |
| 0.380 | 0.690 | 0.386 | 0.542 | 0.695 | 0.081 |